\documentclass[aps,pre,superscriptaddress,preprint]{revtex4-1}
\usepackage{amsmath}
\usepackage{graphicx}
\usepackage{color}
\usepackage{subcaption}
\usepackage{natbib}
\usepackage{bmpsize}
\begin{document}

\title{Flux Jamming, Phase Transitions and Layering in Turbulent Magnetized Plasma}

\author{P.H. Diamond}
\affiliation{University of California, San Diego, La Jolla, CA, 92093, USA}

\author{Y. Kosuga}
\affiliation{Research Institute for Applied Mechanics, Kyushu University, Fukuoka, 816-8580, Japan}

\author{P.L. Guillon}
\affiliation{Laboratoire de Physique des Plasmas, Ecole Polytechnique, Palaiseau, 91128, France}

\author{{\"O}.D. G{\"u}rcan}
\affiliation{Laboratoire de Physique des Plasmas, Ecole Polytechnique, Palaiseau, 91128, France}

\begin{abstract}

This paper discusses transport barrier formation and layering as consequences of jam formation. Extensive use is made of analogies with the theory of traffic flow in one dimension. The relation of flux jamming to motility induced phase separation (MIPS) is explained. Two routes to heat flux jamming are identified. The first is due to a rollover in the heat flux-pulse size relation, i.e. $dQ_T(\delta T)/d\delta T<0$, and is similar to the condition of flux-gradient bistability. The second occurs when the delay time between pulse and heat flux exceeds a critical value. This does not require bistability and tends to occur near marginality. This analysis yields an estimate of the answer to the eternal question of 'how near is "near"?'. Staircase development is shown to follow jamiton train formation. The relation of jamming of avalanches to phase transitions in drift wave-zonal flow turbulence is elucidated. The formation of outward propagating blob trains and inward propagating void trains is demonstrated. The important role of turbulence spreading is identified.

\end{abstract}

\maketitle

\section{Introduction}

Jamming is an obvious means for inducing inhomogeneity in transport and mixing. Jamming leads naturally to phase separation - i.e. a flow which jams locally then breaks up into separate domains, bounded by bottlenecks. This, in turn, triggers the formation of layers. Indeed, a traffic flow with some set of internal jamming sites will have the structured pattern of a number of regions of free flow, bounded by a similar number of bottlenecks, where the flow is reduced to a trickle. Sharp density gradients develop in the jams. This configuration is reminiscent of a staircase profile, where 'steps' - regions of uninhibited transport flux - are located between interspersed 'jumps' or transport barriers, where the flux is reduced locally. Thus, it is appealing to approach the problem of layering or staircase formation from the perspective of \textit{flux jamming}. In particular, a staircased temperature profile can emerge from a train of heat flux jams.

In this paper, we discuss the formation of layers in avalanching drift wave turbulence in plasma due to heat flux jams. Extensive analogies to the theory of 1D traffic flow are utilized\cite{Whitham,LighthillWhitham,Richards,Reflection}. Interestingly, we show that there are at least \textit{two} routes to jamming. The first mechanism is analogous to that for the formation of a backward shock in traffic flow, i.e. $d(\rho V)/d\rho<0$, rather like what happens upon entering a bottleneck (Table \ref{Table:JamComparison}). This condition also is similar to that for motility induced phase separation (MIPS)\cite{MIPS_Cates} which occurs when scattered particles tend to slow down in regions of high density. We show that the analogous condition for jams in an avalanching heat flux is $dQ_T(\delta T)/d\delta T<0$. Here $\delta T$ is the turbulent heat flux. $dQ_T(\delta T)/d\delta T<0$ can result from transport feedback loops which lower the turbulent flux. $E\times B$ shearing - with $E_r$ linked to $\delta T$ by radial force balance\cite{BDT,ExBBurrell} - is one common example of such a feedback loop. We show that such feedback loops are entirely consistent with the required structure of $Q_T(\delta T)$ - i.e. satisfying the condition of joint reflection symmetry, etc. In the actual bottleneck region, the heat flux is carried by the residual diffusion. This mechanism of jam formation resembles barrier formation due to flux bistability, now applied to avalanche pulses. The condition for jam formation is calculated.

\begin{table}[!h]
\caption{Models for jam - correspondence.}
\label{Table:JamComparison}
\centering
\begin{tabular}{lll}
\hline
& Traffic flow & Heat avalanche\\
\hline\hline
Quantify of interest& Car density $\rho$ & Temperature deviation $\delta T$ \\
\hline
Evolution & Continuity  eq. & Heat Balance eq.\\
\hline
Flux & Kinematic flux, $V(\rho)$ & Heat Flux via JRS, $Q_0(\delta T)$  \\
\hline
Roll over condition & $d(\rho V(\rho))/d\rho<0$ & $dQ(\delta T)/d\delta T<0$ \\
\hline
\end{tabular}
\end{table}

A second jamming mechanism involves the interplay of relaxation and diffusion (Table \ref{Table:JamComparison2}). In the 1D traffic problem, the flow is dynamic but tends to relax to $V_0(x)$ on time $\tau_R$. Thus, $\tau_R$ may be thought of as a driver's response time. Similarly, the diffusion plays the role of anticipation. Note that $(D/\tau_R)^{1/2}$ defines a speed. Clustering instability - jamming - occurs for $\tau_R>\nu/(V_0-c_0)^2$ (here $V_0=V(\rho_0)$, $c_0=\rho_0V_0'+V_0$). Thus long reaction time - as for drunk drivers - favors jams. Interestingly, no flavor of bistability enters here - $d(\rho V)/d\rho<0$ is not required.

\begin{table}[!h]
\caption{Dynamic model for jam}
\label{Table:JamComparison2}
\centering
\begin{tabular}{lll}
\hline
& Traffic flow & Heat avalanche\\
\hline\hline
Quantify of interest& Car density $\rho$ & Temperature deviation $\delta T$ \\
\hline
Evolution & Continuity  eq. & Heat Balance eq.\\
\hline
Flux & Traffic speed, $V$ & Instantaneous Heat Flux, $Q$\\
\hline
Mean Flux & Kinematic flux, $V_0(\rho)$ & Heat Flux via JRS, $Q_0(\delta T)$  \\
\hline
Relaxation Time & Drivers' response $\tau_d$ & Turbulence mixing time $\tau_d$ \\
\hline
\end{tabular}
\end{table}

In this vein, then, we consider an evolving heat flux, which undergoes turbulent mixing and which relaxes to its mean field value. Thus, Ficks Law is replaced by a Guyer-Krumhansal type constitutive relation\cite{GuyerKrumhansl,AvaJamSecond}. Here, $\tau_R$ is the turbulent mixing time of the heat flux. An analogous analysis yields the jamming condition $\tau_R>\chi_{neo}/c_0^2$, where $\chi_{neo}$ is the residual neoclassical heat conductivity and $c_0$ is the avalanche speed\cite{AvaJamLett,AvaJamFull}. Long $\tau_R$ means long mixing time, which naturally occurs 'near marginality'. The question is: how near?. Taking the ansatz $1/\tau_R\sim1/\tau_0[R/L_T-R/L_{Tc}]^\mu$, we show that the jam condition is $1>(\chi_{neo}/c_0^2)(1/\tau_0)[R/L_T-R/L_{Tc}]^\mu$. Here $\mu$ is the gradient 'stiffness exponent' and $\tau_0$ is $\tau_c$ evaluated for mixing length values - i.e. as a normalization constant. The result above answers, at last, the age old question of what exactly the oft-heard phrase 'near to marginal' \textit{actually means} in the context of jams and layering. As with traffic flow, jams form independent of the $dQ_T/d\delta T<0$ condition.

To connect to layering we report on studies of trains of jamming pulses, or jamitons\cite{Jamiton,AvaJamNum}. We show that these trains evolve sequentially. An extended jamiton train produces a sequence of temperature modulations. When superimposed on a smooth background, a temperature staircase profile results.

To further elucidate avalanche physics we also present a study of particle flux avalanching. This system manifests avalanching due to localized relaxation as an outgoing train of blobs ($\tilde{n}>0$) and an ingoing train of voids ($\tilde{n}<0$). Signatures of criticality phenomena appear near the transition from eddy to zonal flow states. Intensity field evolution is consistent with the 1D model of avalanching/spreading discussed here.

The remainder of this paper is organized as follows. Section II discusses avalanching turbulence and how inhomogeneous mixing occurs in it. Section III develops the theory of heat flux jams in extended kinematic wave theory. Section IV discusses jams in dynamics, with heat conduction and relaxation. Section V discusses relevant aspects of particle flux avalanching in dissipative drift wave turbulence. Section VI contains the discussion and conclusion.

\begin{figure}[!h]
\centering\includegraphics[width=0.8\textwidth]{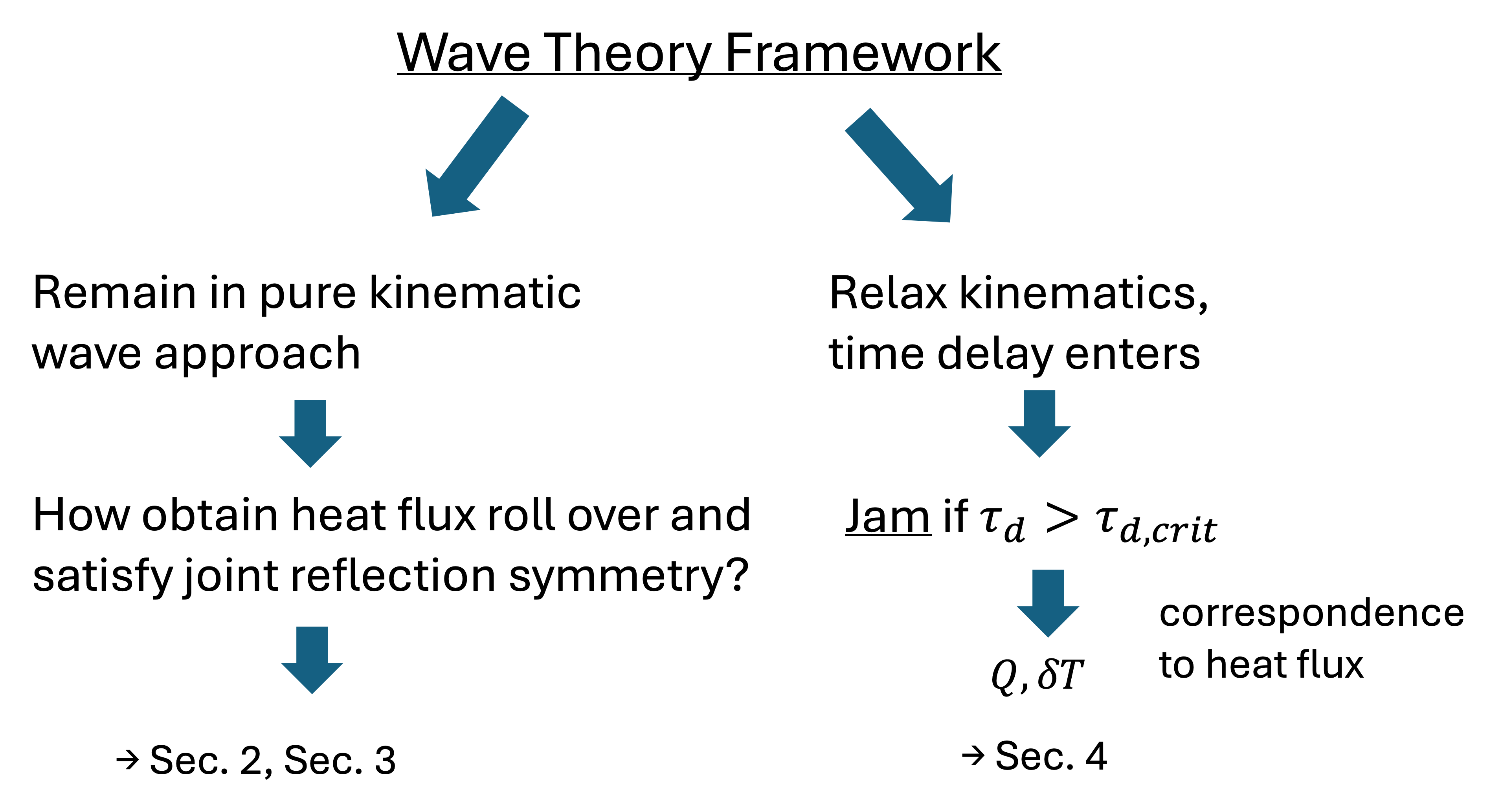}
\caption{Approaches to Jam}
\label{Fig:Flow}
\end{figure}

\section{Avalanching Turbulence - How Is Inhomogeneous Mixing Achieved?}
\begin{figure}[!h]
\centering\includegraphics[width=0.5\textwidth]{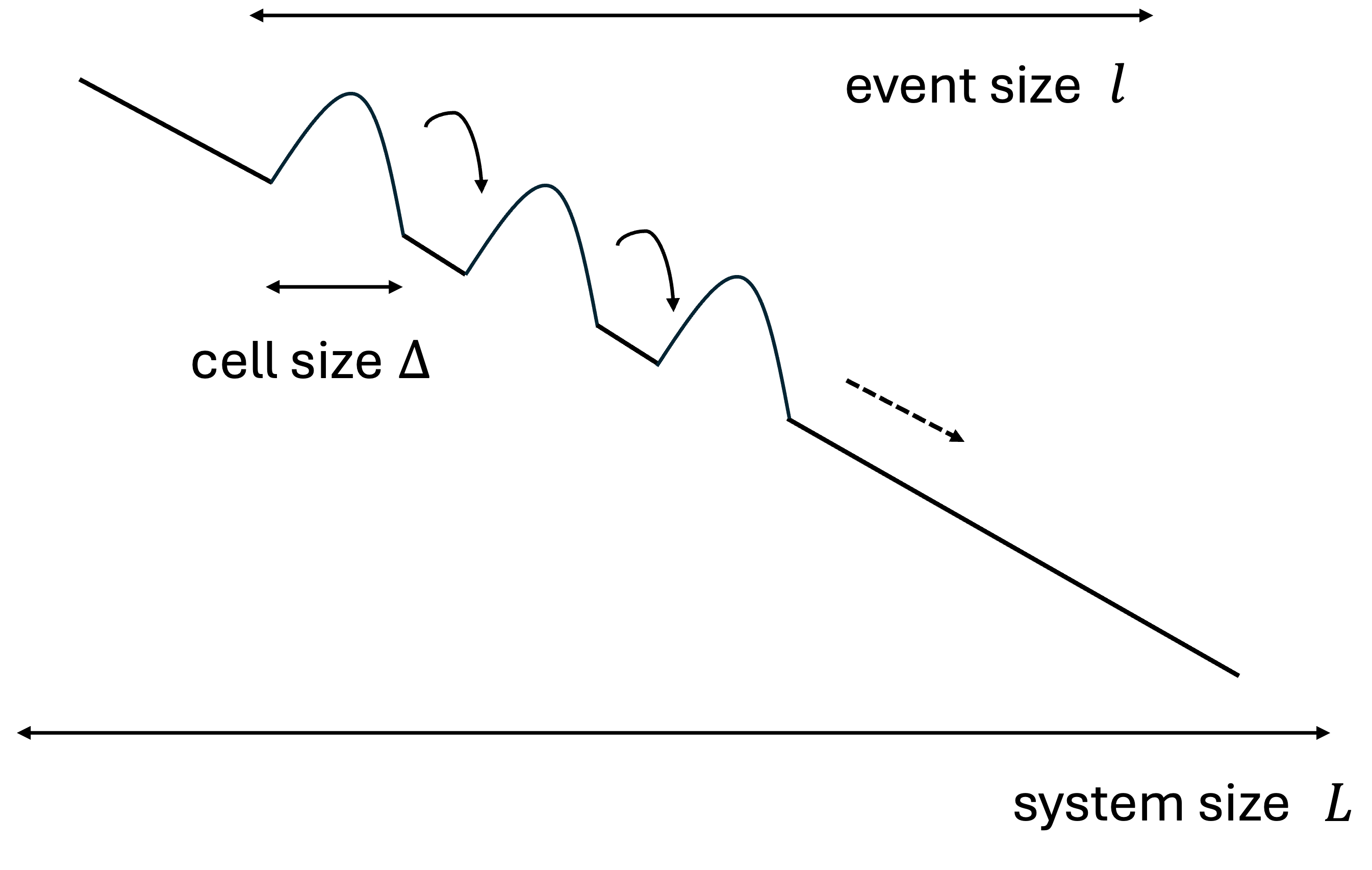}
\caption{A cartoon for an avalanche event. Also shown are relevant length scales, i.e. cell size $\Delta$, event size $l$ and the system size $L$.}
\label{Fig:Size}
\end{figure}

Confined plasma turbulence manifests avalanches, which are a statistical distribution of intermittent, radially extended transport events\cite{BTW,AvalancheDIII-D,AvalancheKSTAR,AvalancheJT60,AvalancheHeliotron,AvalanhceLAPD,HahmDiamondReview}. Avalanches are propagating fronts of correlated overturnings or gradient relaxation events, accompanied by a pulse of turbulence\cite{Townsend,StructureSpreading,VoidSpreading,SpreadingDIII-D}. Avalanches arise naturally, since fluctuations are spatially localized (pinned) with cell size $\Delta\ll L$, where $L$ is the system size (Fig.\ref{Fig:Size}). Thus $\Delta/L\ll1$. Local relaxation occurs when a critical local profile gradient is exceeded, so the cell is excited. Interaction with neighboring cells then leads to avalanching. The production ratio
\begin{equation}
R_p=\frac{\Delta\langle\tilde{v}_r\tilde{n}\tilde{n}\rangle}{-\int dr\left[\langle\tilde{v}_r\tilde{n}\rangle\frac{\partial\langle n\rangle}{\partial r}\right]}
\end{equation}
- the ratio of the increment in intensity flux across a finite region to the integrated local production in that region - is a practical measure of the strength of cell-to-cell interaction effects in turbulence excitation. $R_p$ has been measured\cite{SpreadingRate,SpreadingRatio} and frequently exceeds unity, indicating interaction dominates over local production. We note $R_p$ is related to, but not identical to, Kubo number. Detailed simulations\cite{AvalancheRI,AvalancheITGFluid,AvalancheFluidRev,AvalancheGKFrench,AvalancheGKJapan} of plasma avalanching observed $\sim 1/f$ spectra, and a Zipf's law, $P(l)\sim1/l$, probability distribution of avalanche size $l$. Avalanches tend to occur near marginality or criticality, when fluctuation Kubo number $Ku\sim\tilde{v}_r\tau_m/\Delta_c\gtrsim1$. Here $\tilde{v}$ is the rms of the radial $E\times B$ fluctuation velocity, $\tau_m$ is the memory time, and $\Delta_c$ is the correlation scale. The latter basically is the cell size $\Delta_c\sim\Delta$. $Ku>1$ arises for long $\tau_m$, indicative of extended memory. Here 'long' means $\tau_m>\tau_c$, where $\tau_c$ is the auto-correlation time of an individual cell. Long memory is a result of the collective interaction of many cells in an avalanche.

\begin{figure}[!h]
\centering\includegraphics[width=0.5\textwidth]{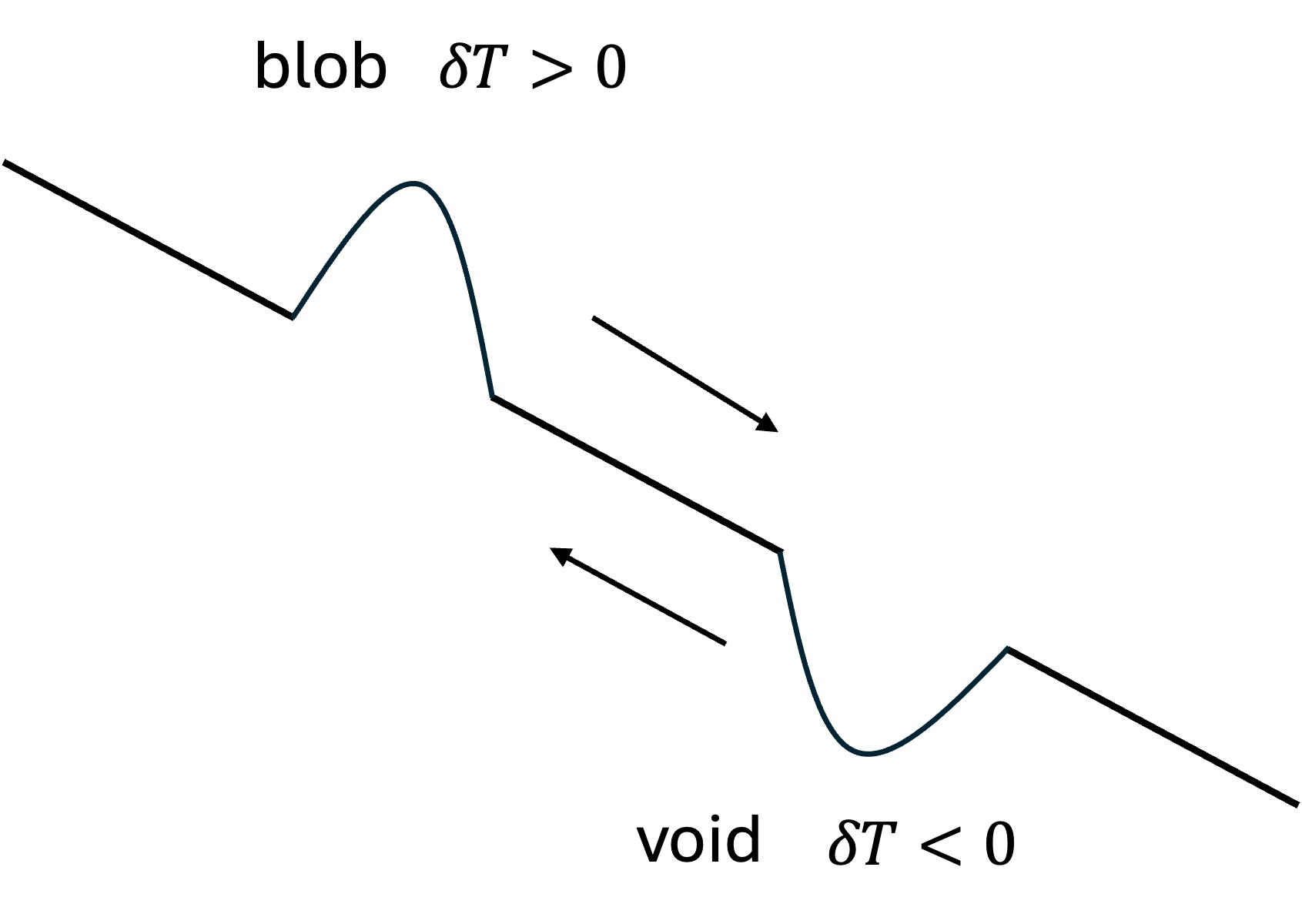}
\caption{Outgoing blob and incoming void}
\label{Fig:BlobVoid}
\end{figure}

\begin{figure}[!h]
\centering\includegraphics[width=0.5\textwidth]{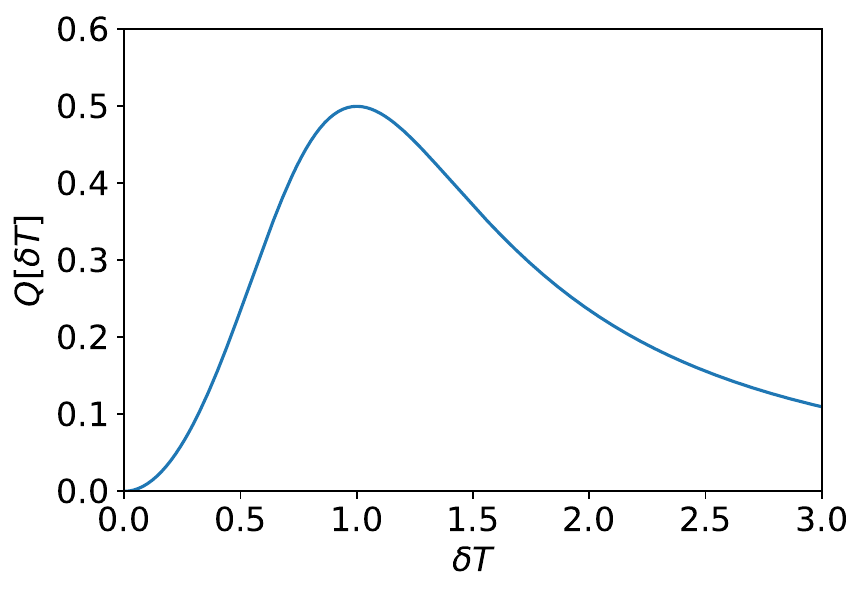}
\caption{An example of \textit{u-shaped} flux. For large $\delta T$, the flux decreases, i.e. $dQ/d\delta T<0$. The rollover is a route for various relevant phenomena, such as transport bifurcation, jamming, etc, as explained in detail in the text.}
\label{Fig:Ushape}
\end{figure}

A tractable, coarse grained model of avalanching turbulence is of great interest. The simplest such model describes the evolution of large scale fluctuations $\delta T$ about the (dynamic) mean profile according to:
\begin{equation}
\frac{\partial}{\partial t}\delta T+\partial_xQ[\delta T]-\chi_0\partial_x^2\delta T=\tilde{s}
\label{Eq:HeatBalance}
\end{equation}
Here $\delta T$ refers to the large scale excursion from the dynamic mean profile. 'Large scale' indicates scales $l>\Delta$. Note that Eq.(\ref{Eq:HeatBalance}) does not determine the mean profile. This is a minimal 1D model of heat avalanche dynamics. Such avalanches propagate in the radial direction, denoted by $x$. The effect of avalanches is to dynamically corrugate the mean profile. $Q[\delta T]$ is the radial ($\hat{x}$) heat flux which evolves $\delta T$ in time, $\chi_0$ is ambient heat conduction (collisional or neoclassical) and $\tilde{s}$ is noise. Here $\delta T$ is a conserved order parameter. The most general form of $Q[\delta T]$ is determined by requiring its consistency with the principle of joint reflection symmetry (JRS)\cite{SurfaceGrowth,Avalanche}, i.e. invariance under dual transformations of $\delta T\to -\delta T$ and $x\to-x$. This ensures that $Q[\delta T]$ respects the condition that excesses or 'blobs' ($\delta T$) move down-gradient while deficits or 'voids' ($\delta T<0$) move up-gradient (Fig.\ref{Fig:BlobVoid}). Note this approach to the structure of $Q[\delta T]$ follows the spirit of Landau-Ginzburg theory - namely the use of symmetry constraints to deduce the set of possible $Q[\delta T]$. It should be contrasted with the usual perturbative approaches. The smoothest form of $Q[\delta T]$ allowed by JRS is usually taken as:
\begin{subequations}
\begin{equation}
Q[\delta T]=\frac{\alpha}{2}(\delta T)^2-D\frac{\partial}{\partial x}\delta T
\label{Eq:BurgersFlux}
\end{equation}
so $\delta T$ evolves according to 
\begin{equation}
\frac{\partial}{\partial t}\delta T+\alpha\delta T\partial_x\delta T-\chi_0\partial_x^2\delta T=\tilde{s}
\label{Eq:Burgers}
\end{equation}
Here $\alpha$ is a coefficient to be specified and $D$ has been absorbed into $\chi_0$. Equation (\ref{Eq:Burgers}) is the familiar noisy Burgers equation, which describes 1D avalanches as shock turbulence.
\end{subequations}
More generally, it is natural to classify contributions to $Q[\delta T]$ according to their sensitivity to scale, with more sensitive contributions, containing higher order derivatives, being less significant on larger scales. A more general form of $Q[\delta T]$ is shown in Eq. (\ref{Eq:JRS}). Here $f$ is a function to be specified.
\begin{equation}
Q[\delta T]=\frac{\alpha}{2}(\delta T)^2/f\left[\beta(\delta T)^{2n}+\gamma\left(\frac{\partial\delta T}{\partial x}\right)^{2m}+...\right]-D\frac{\partial}{\partial x}\delta T
\label{Eq:JRS}
\end{equation}
Equation (\ref{Eq:JRS}) is entirely consistent with JRS and manifests the critical property that $Q[\delta T]$ can roll over and decay for sufficiently large $\delta T$, $\partial\delta T/\partial x$ etc. This allows the triggering of jams and transport barriers when critical values of $\delta T$, $\partial\delta T/\partial x$, etc. are exceeded - i.e. for sufficiently large and/or steep heat avalanches. For simplicity, we take $f=c(1+\beta(\delta T)^{2n})$ and absorb the factor $c$ into $\alpha$, hereafter. Thus, we will examine how $\delta T$ evolves according to 
\begin{subequations}
\begin{equation}
Q[\delta T]=\frac{\alpha/2(\delta T)^2}{1+\beta(\delta T)^{2n}}-D\frac{\partial}{\partial x}\delta T
\label{Eq:MIPSFlux}
\end{equation}
so then $\delta T$ evolves according to:
\begin{equation}
\frac{\partial}{\partial t}\delta T+\frac{\partial}{\partial x}\left[\frac{\alpha/2(\delta T)^2}{1+\beta(\delta T)^{2n}}\right]-\chi_0\partial_x^2\delta T=\tilde{s}
\label{Eq:MIPS}
\end{equation}
For sufficiently small $\delta T$, Eq. (\ref{Eq:MIPS}) effectively reduces to the Burgers model, while for sufficiently large $\delta T$, $Q[\delta T]$ \textit{decreases} for increasing $\delta T$ (i.e. $dQ/d\delta T<0 $), indicative, as we shall see, of possible heat flux jam formation (Fig.\ref{Fig:Ushape}). 
\end{subequations}

To address the natural question of the \textit{physics} of the rollover and decay of $Q$!?, we recall the now well-established concept of transport barrier formation by bifurcation between roots of a bistable flux function\cite{ScurveHinton,Bistability,BistabilityHW,BistabilityScale}. A simple example is
\begin{subequations}
\begin{equation}
Q=-\frac{\chi_T}{1+\beta\langle V_E'\rangle^2}\frac{\partial\langle T\rangle}{\partial x}-\chi_{neo}\frac{\partial\langle T\rangle}{\partial x}
\label{Eq:Hinton}
\end{equation}
where $\langle V_E'\rangle=\partial_r(-c\langle E_r\rangle/B)$ is the sheared poloidal $E\times B$ velocity determined by radial force balance
\begin{equation}
\langle E_r\rangle=\left\langle\frac{\partial_rp}{n}\right\rangle-\langle\mathbf{V}\rangle\times\mathbf{B}
\end{equation}
The physics of Eq.(\ref{Eq:Hinton}) is that $E\times B$ shear tends to quench transport by eddy tilting and stretching\cite{BDT,ExBBurrell}. Note that the diamagnetic $E\times B$ velocity ($\sim\partial_rp$) couples the $E\times B$ shear suppression factor to the turbulent heat flux, creating a negative feedback loop of the form: steepen $\partial_r\langle T\rangle$ $\to$ increase $\langle V'_E\rangle$ $\to$ reduce turbulent heat flux $\to$ steepen $\partial_r\langle T\rangle$ $\to$ etc. This results in a transport bifurcation from a turbulent heat flux ($\sim-\chi_T\partial\langle T\rangle/\partial x$) to the residual neoclassical (i.e. collisional) heat flux ($-\chi_{neo}\partial\langle T\rangle/\partial x$), along with a concomitant steepening of the temperature gradient. Such regions of steepeend temperature gradient are frequently referred to as transport barriers. Note also that a heat pulse - i.e. a sudden increase in the heat flux - can trigger the feedback loop discussed above by an impulsive increase in the local temperature gradient. Indeed, it is well known that a heat pulse can trigger the L$\to$H transition\cite{HmodeWagner,II,HmodeWagnerReview,LHPhilTrans}. It is this property of self-suppression of the heat flux that underpins the consideration of forms for $Q[\delta T]$ which lead to jam formation, i.e. a $Q[\delta T]$ which tends to rollover and decay for large $\delta T$.
\end{subequations}

\section{Heat Flux Jams in Extended Kinematic Waves}
\begin{figure}[!h]
\centering
\begin{subfigure}[b]{0.45\textwidth}
\includegraphics[width=\textwidth]{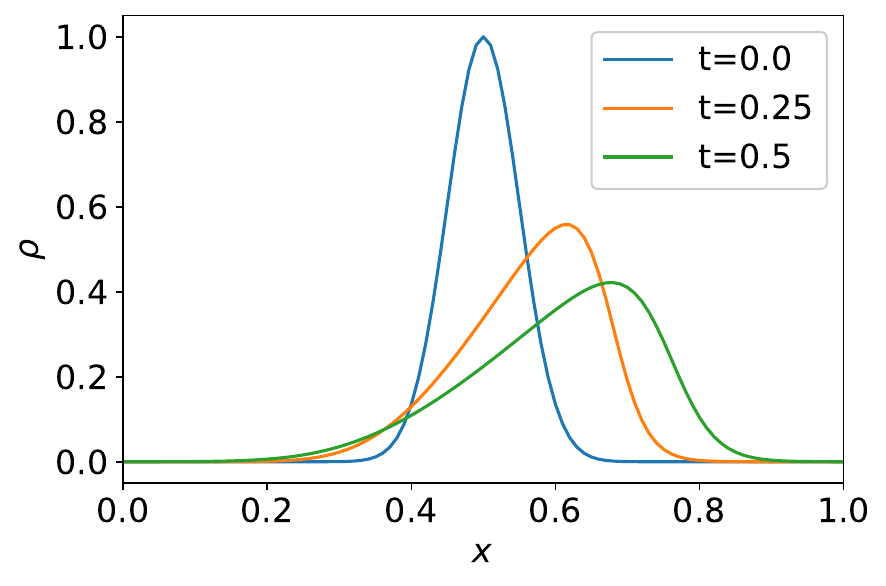}
\caption{$c(\rho)=\rho$, $D=0.01$}
\end{subfigure}
\centering
\begin{subfigure}[b]{0.45\textwidth}
\includegraphics[width=\textwidth]{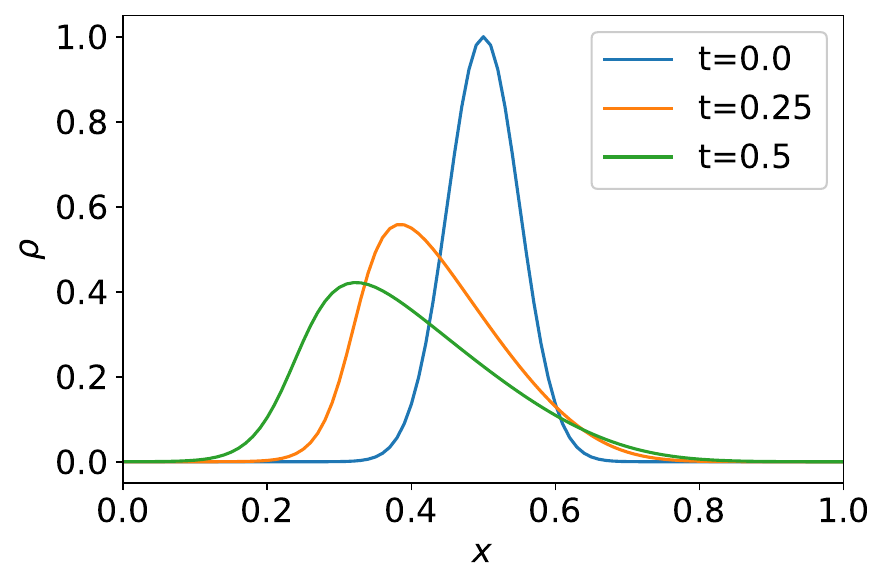}
\caption{$c(\rho)=-\rho$, $D=0.01$}
\end{subfigure}
\caption{Forward and backward shocks}
\label{Fig:Shocks}
\end{figure}

Readers will no doubt recognize Eq.(\ref{Eq:MIPS}) as a particular incarnation (for $\tilde{s}=0$) of the kinematic wave equation of Lighthill and Whitham\cite{Whitham,LighthillWhitham}, and Richards\cite{Richards}
\begin{equation}
\frac{\partial\rho}{\partial t}+\frac{\partial}{\partial x}\left(\rho V(\rho)\right)-D\frac{\partial^2\rho}{\partial x^2}=0
\label{Eq:KinematicWave}
\end{equation}
This is studied extensively in the context of traffic flow\cite{Reflection}. Here kinematic refers to the fluid velocity being taken as a given function of density ($v=V(\rho)$), rather than evolving dynamically. In Eq.(\ref{Eq:MIPS}), $Q[\delta T]$ similarly is a given function of $\delta T$, alone. Thus, kinematic wave theory is relevant to both cases. Early studies deduced $V(\rho)$ empirically, and found $V(\rho)$ to have an inverted u-shaped profile - i.e. increasing at low $\rho$, then ultimately decreasing at large $\rho$. A striking feature of kinematic waves is that they can manifest both forward and backward shocks, for $d(\rho V(\rho))/d\rho>0$ and $d(\rho V(\rho))/d\rho<0$, respectively. This is readily apparent from re-writing Eq.(\ref{Eq:KinematicWave}) as
\begin{equation}
\frac{\partial}{\partial t}\rho+c(\rho)\frac{\partial}{\partial x}\rho-D\frac{\partial^2\rho}{\partial x^2}=0
\end{equation}
where $c(\rho)=d(\rho V(\rho))/d\rho$ is the pattern speed. For $\rho$ such that $c(\rho)>0$, forward shocks, akin to those familiar from gas dynamics, can form. For $\rho$ such that $c(\rho)<0$, the density perturbation propagates in the opposite direction to the flow, leading to the formation of a backward shock. Forward and backward shocks are sketched in Fig.\ref{Fig:Shocks}.

Backward shocks are familiar from the common experience of needing to brake suddenly when encountering a traffic bottle neck. The abrupt change in density at the face of a backward shock implies a transition from high speed flow to low speed flow, otherwise known as a \textit{traffic jam}. Jamming occurs as a consequence of the negative feedback loop implied by the condition that $d(\rho V(\rho))/d\rho<0$, i.e. increment in $\rho$ ($\rho\to\rho+\delta\rho$) $\Rightarrow$ decrement in $V(\rho)$ ($V\to V+\delta V$, where $\delta V<0$) $\Rightarrow$ further increment in $\rho$, etc leading to an effective separation of the flow into 'phases' of normal speed and jams. The similarity of this feedback loop to that of the $E\times B$ shear-induced transport barrier is evident. The normal $\to$ jam phase boundary is the backward-facing shock. \textit{The backward shock may be thought of as a transport barrier}. The analogy between jams and transport barriers will be developed below. We note here that a jam is effectively an expanding barrier, which propagates upstream, while vehicles move downstream, i.e. forward.

The observation that jam formation induces phase separation in the flow brings us to its relation to MIPS - motility induced phase separation, a concept developed by Cates and Tailleur\cite{MIPS_Cates}, and broadly relevant in active matter and fluids, biophysics, etc. MIPS occurs when self-propelled particles (or vehicles) slow down in dense regions. This results in a negative feedback loop of the form: reduced motility $\rightarrow$ accumulation (if the flow speed decreases sufficiently rapidly as $\rho$ increases) $\to$ further slowdown $\to$ clustering. The upshot is the separation into, and coexistence of, dense and dilute regions without an explicit appeal to any difference in interactions, in either group. Interestingly, the criterion for the onset of MIPS for a 1D flow $V(\rho)$ given in Eq. (3) of Cates and Tailleur\cite{MIPS_Cates} is:
\begin{equation}
\frac{V'(\rho_0)}{V(\rho_0)}<-\frac{1}{\rho_0}
\end{equation}
which is manifestly identical to the backward shock condition $d(\rho V(\rho))/d\rho<0$. Thus, traffic jam formation is seen as a realization of MIPS in one dimension. More generally, a close relationship between jamming and MIPS is evident from the above discussion. We add that the underlying character of systems manifesting MIPS follows from the absence of time reversal symmetry in active systems (including traffic flow), so that their steady states need not obey the well known principle of detailed balance (DB). DB states that if phase space is divided into different sectors, then the probability flux from sector A $\to$ sector B equals the reverse probability flux sector B $\to$ sector A. DB is clearly violated by jam formation and, more generally, for non-reciprocal interactions, one of example of which is generation of zonal flows by drift-Rossby wave turbulence and, more generally, transport barrier formation. These two processes have been described by manifestly non-reciprocal predator-prey models.

It is evident, then, that turbulent heat flux jams can occur in the model of Eq.(\ref{Eq:MIPS}) if $dQ/d\delta T<0$ for some interval of $\delta T$. Overall continuity of flux then implies that $d\delta T/dx$ must steepen at the jam locations, in analogy to shocks. Transport into the jams will be regulated by the residual diffusion, just as neoclassical diffusion regulates transport in regimes of strong $E\times B$ shear, such as barriers.

\begin{figure}[!h]
\centering\includegraphics[width=0.5\textwidth]{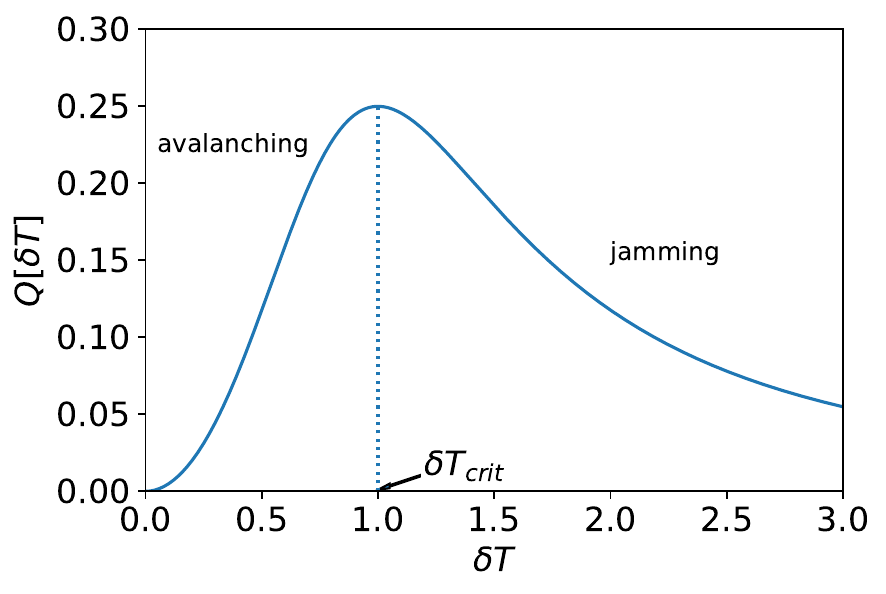}
\caption{The case with $\alpha=1$, $\beta=1$, $n=2$. Branch crossing corresponds to Jam boundary.}
\label{Fig:CritRollover}
\end{figure}

Jams will occur for $dQ/d\delta T<0$, so that pulses which jam occurs on the decreasing branch of the $Q$ vs. $\delta T$ curve. For $Q$ as given by Eq.(\ref{Eq:MIPSFlux}),
\begin{subequations}
\begin{equation}
\frac{dQ}{d\delta T}=\frac{\alpha\delta T}{1+\beta(\delta T)^{2n}}\left[\frac{1-(n-1)\beta\delta T^{2n}}{1+\beta(\delta T)^{2n}}\right]
\end{equation}
so jams occur for $n>1$ and
\begin{equation}
\delta T>\delta T_{crit}=\left(\frac{\beta}{n-1}\right)^{1/2n}
\label{Eq:MIPSCritical}
\end{equation}
Equation (\ref{Eq:MIPSCritical}) identifies the heat pulse jamming threshold - i.e. a critical avalanche size in order to jam and form a barrier. Note that $\delta T_{crit}$ requires $n>1$, so as the flux roll over fast enough with increasing $\delta T$, and that $\delta T$ exceed a factor set by the suppression coefficient $\beta$. Here, jam formation is a consequence of the breaking of rescaling invariance of $Q[\delta T]$, i.e. the appearance of two branches, as shown in Fig.\ref{Fig:CritRollover}. A related question concerns the jam locations. For $\beta$ slowly varying in space, so $\beta=\beta(x)$, it is evident that $\delta T_{crit}$ will track the local maxima in $\beta(x)$, and thus so will the jam locations. Thus an oscillatory $\beta(x)$, as for a quasi-periodic shearing field, should naturally trigger the formation of an array of jams, as in an $E\times B$ staircase\cite{PVStaircase,StaircaseSim,StaircaseSim2,StaircaseExp,StaircasePermeability,StaircaseDIII-D,StaircaseKSTAR,StaircaseHL2A,Ramirez}. This suggests that staircases form when the pattern of heat flux jams 'locks on' to modulations in the shearing field.

\end{subequations}

\section{Jams in Dynamics, with Diffusion and Relaxation}
\begin{figure}[!h]
\centering\includegraphics[width=0.5\textwidth]{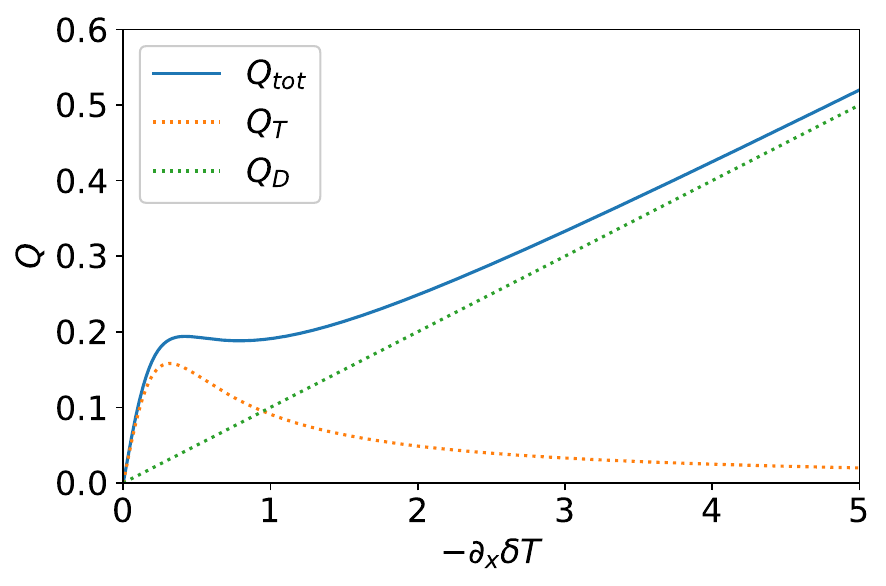}
\caption{An example of S-curve flux-gradient relation}
\label{Fig:SCurve}
\end{figure}

Up till now, we have not strayed from the realm of kinematic wave theory, and an erudite reader would be justified in making the remark "Isn't the supposition that $dQ/d\delta T<0$ (equivalently $d(\rho V/d\rho<0)$) for a range of $\delta T>\delta T_{crit}$ tantamount to presupposing bistability of the flux - a familiar theme?" This is easily seen by observing that combining $Q_{T}(\delta T)$ of the u-shape with residual diffusion, $Q_{D}=-Dd\delta T/dx$, $D$: constant, evokes the familiar S-curve (Fig.\ref{Fig:SCurve}), frequently used to describe turbulent bifurcations. Simply put, the $Q$ vs. $-\delta T'$ s-curve above evolves from $Q_{tot}=Q_{T}+Q_{D}$. Thus, it is not entirely surprising that this model manifests jams, transport bifurcations, and phase separation phenomena. However, relaxing the constraints of kinematic wave theory by considering \textit{dynamics} opens the road to a different, and highly relevant, jamming mechanism. It is to this we now turn.

Dynamic waves loosen the constraints of kinematic wave theory by positing that the flow $V$ relaxes to its kinematic value in a time $\tau_R$. The dynamic traffic flow equations then are:
\begin{subequations}
\begin{align}
&\frac{\partial}{\partial t}\rho+\partial_x(\rho v)=0\\
&\frac{\partial}{\partial t}v+v\partial_xv=-\frac{1}{\tau_R}(v-V(\rho)+\frac{\nu}{\rho}\partial_x\rho)
\label{Eq:TrafficJamSpeed}
\end{align}
\end{subequations}
For short relaxation times, $v\to V(\rho)-\nu\partial_x\rho/\rho$, recovering kinematic wave theory. For long relaxation times, significant deviations of the flow from its kinematic value are possible. $\tau_R$ and $\nu$ model the capacity of drivers for response and anticipation, respectively. Note, too, that $\nu$ and $\tau_R$ taken together define a speed $(\nu/\tau_R)^{1/2}$, thus breaking the rescaling invariance of purely kinematic models. It should be noted that the dynamic equations can be derived as the continuum limit of a discrete car following model. There, $\tau_R$ and $\nu$ emerge from the car-to-car interaction rules, which resemble cellular automata rules. In this sense, the dynamic flow model can be seen as an antecedent of the 'active fluid' concept.

\begin{figure}[!h]
\centering\includegraphics[width=0.5\textwidth]{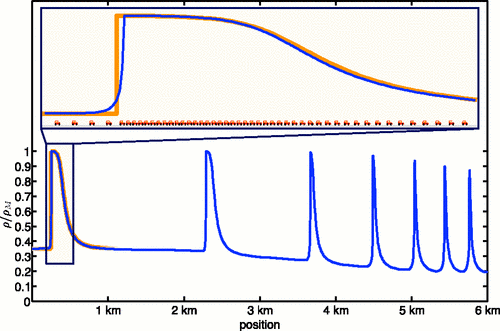}
\caption{Jamiton, a nonlinear wave in traffic flow dynamics model\cite{Jamiton} (Copyright APS)}
\label{Fig:Jamiton}
\end{figure}

The stability of dynamic flows is discussed in detail by Whitham\cite{Whitham}. The linearized theory is useful for its convenience, clarity and simplicity. Writing $\rho=\rho_0+r$, $v=v_0+w$ ($v_0=V(\rho_0)$) and taking $c_0=\rho_0V'(\rho_0)+V(\rho_0)$ ($V'(\rho_0)=dV(\rho)/d\rho|_{\rho_0}$) and eliminating $w$ gives
\begin{equation}
\frac{\partial r}{\partial t}+c_0\frac{\partial r}{\partial x}=\nu\frac{\partial^2r}{\partial x^2}-\tau_R\left(\frac{\partial}{\partial t}+V_0\frac{\partial}{\partial x}\right)^2r
\label{Eq:DensityPerturb}
\end{equation}
This can be further simplified by looking for solutions of the form $f=f(x-c_0t)$, so $\partial_t\cong-c_0\partial x$. Then Eq.(\ref{Eq:DensityPerturb}) reduces to
\begin{equation}
\frac{\partial r}{\partial t}+c_0\frac{\partial r}{\partial x}=\left(\nu-\tau_R(V_0-c_0)^2\right)\frac{\partial^2r}{\partial x^2}
\label{Eq:DensityPerturb2}
\end{equation}
It is apparent that the effective diffusivity in Eq.(\ref{Eq:DensityPerturb2}) goes negative for $\nu<\tau_R(V_0-c_0)^2$, indicating instability. Thus, clustering via negative diffusion will occur for $\nu<\tau_R(V_0-c_0)^2$. Note that this condition is \textit{independent} of the sign of $V'(\rho_0)$ and so is fundamentally different from that discussed in section III, which requires
\begin{equation}
d(\rho V(\rho))/d\rho<0
\label{Eq:JamCondition}
\end{equation} In particular, \textit{bistability} of the traffic flux is \textit{not} required! Obviously, conditions of $\nu/\tau_R<(V_0-c_0)^2$ - i.e. reduced "anticipation" and long "reaction time" favor jamming instability. A moments contemplation of the possible impact of driving while intoxicated will drive home the plausibility of this result. Linear jams and jamming waves have been shown to develop into nonlinear waves, called 'Jamitons' (Fig.\ref{Fig:Jamiton}).

In avalanching heat transport, it is then natural to consider the possibility of departures of the instantaneous heat flux from an expected or mean value, with the supposition that the former will tend to relax toward the latter. At the simplest level, this would, say, replace the familiar Fick's Law
\begin{subequations}
\begin{equation}
Q=-\chi\nabla T
\end{equation}
with a Guyer-Krumhansl equation\cite{GuyerKrumhansl} of the form
\begin{equation}
-\tau\frac{\partial Q}{\partial t}-\chi\nabla T=Q
\end{equation}
\end{subequations}
The latter may be derived systematically via moments of the fluctuation entropy (i.e. intensity of phase space density fluctuations) equation. The response time $\tau_R$ is set by the turbulent decorrelation rate. The systematic calculation\cite{AvaJamSecond,AvaJamFull} gives
\begin{equation}
\frac{\partial Q}{\partial t}+\partial_x\Gamma_Q(\mathcal{E})=-k_y^2D_y(\mathcal{E})(Q-\chi(\mathcal{E})\partial_x\langle T\rangle)
\label{Eq:DelayPhasetrophy}
\end{equation}
Here $\mathcal{E}$ is turbulence energy, $k_y^2D_y(\mathcal{E})\sim1/\tau_R$, and $\Gamma_Q$ is the transport of heat flux, due to turbulence spreading. $\mathcal{E}\sim\delta T$, i.e. avalanches necessarily involve turbulence and turbulence spreading. A simplified form of Eq.(\ref{Eq:DelayPhasetrophy}) which neglects spreading is
\begin{subequations}
\begin{equation}
\frac{\partial}{\partial t}Q=-\frac{1}{\tau_R}(Q-Q_0(\delta T))
\label{Eq:Qevo}
\end{equation}
where we have now replaced the Fickian flux with that determined by symmetry (JRS) considerations alone, i.e.
\begin{equation}
Q_0(\delta T)=\frac{\alpha}{2}(\delta T)^2-\chi_2\partial_x\delta T+\chi_4\partial_x^3\delta T
\label{Eq:Q0}
\end{equation}
Here $Q_0(\delta T)$ is effectively equivalent to $Q(\delta T)$ given by Eq.\ref{Eq:JRS}. $\tau_R$ is the relaxation time. $\chi_2$ is equivalent to $\chi_{neo}$, the neoclassical heat conductivity, and a hyper-diffusive contribution is added for reasons of numerical stability. Since jam formation does not required $dQ/d\delta T<0$, we neglect contributions to $Q_0$ which might induce kinematic wave jams, for simplicity.
\end{subequations}

The inquiring reader will likely ask about the underlying microphysics of the relaxation time $\tau_R$ which appears in the flux evolution equation
\begin{equation}
\frac{\partial}{\partial t}Q=-\frac{1}{\tau_R}(Q-Q_0[\delta T])
\end{equation}
Since $Q$ is quadratic in fluctuation level, the fundamental equation is that for the two point phase density fluctuation correlation $\langle\tilde{f}(1)\tilde{f}(2)\rangle$\cite{Dupree,DII,BiglariTIM,KosugaTIMZF,KosugaTIM}. This evolves according to a 2-point equation constructed from the nonlinear gyrokinetic equation. For brevity, we give the result for drift kinetics, neglecting parallel acceleration
\begin{subequations}
\begin{align}
\nonumber
&\frac{\partial}{\partial t}\langle\tilde{f}(1)\tilde{f}(2)\rangle+(v_{\parallel1}\nabla_{\parallel1}+v_{\parallel2}\nabla_{\parallel2})\langle\tilde{f}(1)\tilde{f}(2)\rangle\\
&+P_1\langle f_1\rangle+P_2\langle f_2\rangle+\nabla_1\cdot\mathbf{\Gamma}_{Q_1}+\nabla_2\cdot\mathbf{\Gamma}_{Q_2}=0
\end{align}
where
\begin{align}
\mathbf{\Gamma}_{Q_1}=\langle\tilde{\mathbf{v}}_{E\times B}\tilde{f}(1)\tilde{f}(2)\rangle
\end{align}
and
\begin{align}
P=\langle\tilde{\mathbf{v}}_{E\times B}\tilde{f}\rangle\cdot\nabla
\end{align}
Here $P$ accounts for evolution of $\langle\tilde{f}(1)\tilde{f}(2)\rangle$ driven by mean profile gradients and $\mathbf{\Gamma}_{Q}$ accounts for the turbulent mixing of correlations. $\nabla\cdot\mathbf{\Gamma}_{Q}$ will tend to mix and relax $\langle\tilde{f}(1)\tilde{f}(2)\rangle$, in response to mean gradient drive. An estimate of the rate of mixing is simply the decorrelation rate for $E\times B$ scattering, so $1/\tau_R\sim1/\tau_c$ where $\tau_c$ is a turbulence correlation time.
\end{subequations}

We see that the notion of a relaxation time for the flux response has a physical basis in the effect of $E\times B$ scattering on the evolution of 2-point correlation. We emphasize, though, that $\tau_c\sim\tau_R$ is a simple estimate, and that more accurate calculations of $\tau_R$ should be performed, via systematic renormalization of the 2-point equation\cite{BG}.

The key novel feature of this formulation is the relaxation time $\tau_R$, which sets the rate of relaxation of the flux to its 'mean field' value\cite{AvaJamLett}. Note $\tau_R$ introduces an effective inertia to the response of the flux to avalanche perturbations $\delta T$. This is dramatically illustrated by combining Eqs.(\ref{Eq:Qevo}) and (\ref{Eq:Q0}) and the continuity equation for $\delta T$ (Eq.(\ref{Eq:HeatBalance})) to obtain:
 \begin{align}
\frac{\partial}{\partial t}\delta T+\tau_R\frac{\partial^2}{\partial t^2}\delta T+\alpha\delta T\partial_x\delta T=\chi_2\partial_x^2\delta T-\chi_4\partial_x^4\delta T+\tilde{s}
\label{Eq:NLtelegraph}
\end{align}
Here $\tilde{s}$ is noise. The Burgers equation has now morphed into a nonlinear telegraph equation. Of course, in the limit of short $\tau_R$ and long wavelength (i.e. $\tau_R\to0$, $\partial_x\to0$), the Burgers model is recovered. And as is to be expected, interesting phenomena such as jamming, occur for finite $\tau_R$.

\begin{figure}[!h]
\centering\includegraphics[width=0.5\textwidth]{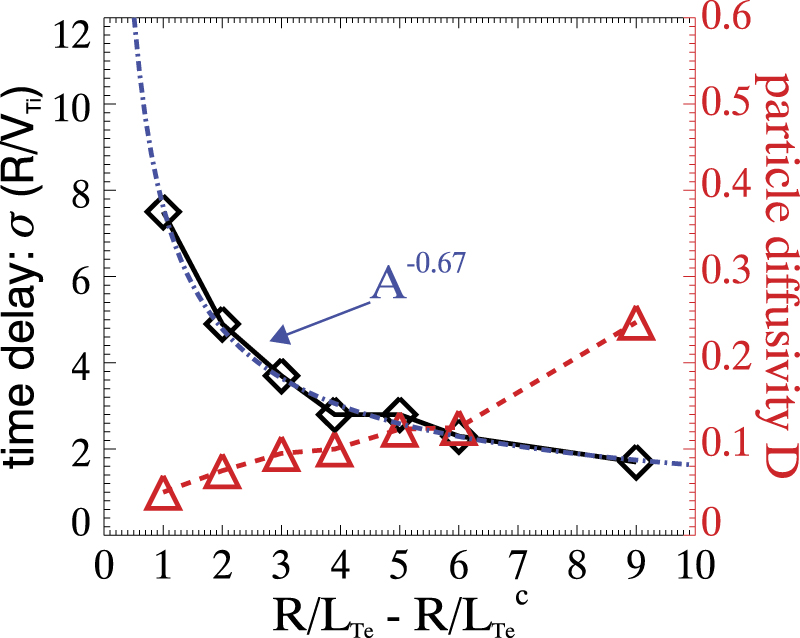}
\caption{Behavior of the response time close to the marginality\cite{GKdelay} (Copyright IOP)}
\label{Fig:ResponseTime}
\end{figure}

Basic properties of the jams can be gleaned from analysis of the 'nonlinear telegraph equation', i.e. Eq.(\ref{Eq:NLtelegraph}). The key novel feature in that is the $\tau_R\partial_t^2\delta T$ term - due to the relaxation time of the heat flux induced by turbulent mixing. Linearizing Eq.(\ref{Eq:NLtelegraph}) according to $\delta T=\delta T_0+\widetilde{\delta T}$, and taking $c_0=\alpha\delta T_0$ to be the unperturbed avalanche speed, we obtain
\begin{subequations}
\begin{equation}
\partial_t\widetilde{\delta T}+c_0\partial_x\widetilde{\delta T}+\tau_R\partial_t^2\widetilde{\delta T}=\chi_2\partial_x^2\widetilde{\delta T}-\chi_4\widetilde{\delta T}
\label{Eq:NLtelegraphLinearized}
\end{equation}
If we look for solutions of the form $\widetilde{\delta T}=f(x-c_0t,\tau)$, where 'fast' variation follows $x-c_0t$ and $\tau$ accounts for slow variation. Then Eq.(\ref{Eq:NLtelegraphLinearized}) reduces to
\begin{equation}
\frac{\partial}{\partial \tau}f=(\chi_2-\tau_Rc_0^2)\partial_x^2f-\chi_4\partial_x^4f
\label{Eq:NLtelegraphLinearizedMoving}
\end{equation}
\end{subequations}
Equation (\ref{Eq:NLtelegraphLinearizedMoving}) is a diffusion equation, with hyper-diffusive correction to control small scales. It is apparent that the sign of the diffusivity is set by that of $\chi_2-\tau_Rc_0^2=\chi_{neo}-\tau_R\alpha^2\delta T_0^2$, so negative diffusion instability occurs for $c_0^2>\chi_{neo}/\tau_R$. This resembles Eq.(\ref{Eq:JamCondition}) - jams occur for long relaxation times, where 'long' means $\tau_R>\chi_{neo}/c_0^2$, or equivalently $c_0^2>\chi_{neo}/\tau_R$ - i.e. \textit{the avalanche speed must exceed the speed of diffusion in a relaxation time.} Since $1/\tau_R\sim1/\tau_c$, the turbulent decorrelation rate, and recalling $1/\tau_c\sim k_\perp^2D(\mathcal{E})\sim \mathcal{E}^\sigma$, where $\sigma>0$, we expect $\tau_R$ will diverge approaching the marginal stability point, i.e. $1/\tau_c\sim(R/L_T-R/L_{Tc})^\mu/\tau_0$ where $0<\mu<1$ and $\tau_0$ is a scale factor. To answer the eternal question of 'how near is "near to marginality"', the stability condition above gives
\begin{equation}
\frac{c_0^2}{\chi_{neo}}>\frac{1}{\tau_0}\left(\frac{R}{L_T}-\frac{R}{L_{Tc}}\right)^\mu
\end{equation}
Thus we see that conditions of weak residual (neoclassical) diffusivity and larger avalanche speed (i.e. larger pulses) favor longer $\tau_R$, and thus jam formation. A similar dependence is also reported from gyrokinetic simulations\cite{GKdelay} (Fig.\ref{Fig:ResponseTime}).

A systematic analysis\cite{AvaJamLett,AvaJamFull} yields the instability frequency and growth rate, and the threshold value for $\tau_R$:
\begin{equation}
\tau_R>\frac{\chi_2}{c_0^2}\left(1+\frac{\chi_4k^2}{\chi_2}\right)
\end{equation}
This recovers the heuristic result with a small correction. The wavenumber for maximal growth then is 
\begin{equation}
k_{max}^2\simeq\frac{\chi_{neo}}{\chi_4}\sqrt\frac{\chi_4c_0^2}{4\chi_{neo}^3}\sim\frac{\alpha\delta T_0}{2\sqrt{\chi_{neo}\chi_4}}
\end{equation}
The scale of the most unstable fluctuation is then $\Delta_{max}^2\sim k_{max}^{-2}$. Maximal growth is $\gamma_{max}\sim c_0/(2l_{diff})$, $l_{diff}\sim(\chi_{neo}\tau_R)^{1/2}$. Thus, the maximal growth rate of jamming is set by the velocity of avalanche propagation and the neoclassical diffusion length in a relaxation time.

\begin{figure}[!h]
\centering\includegraphics[width=0.5\textwidth]{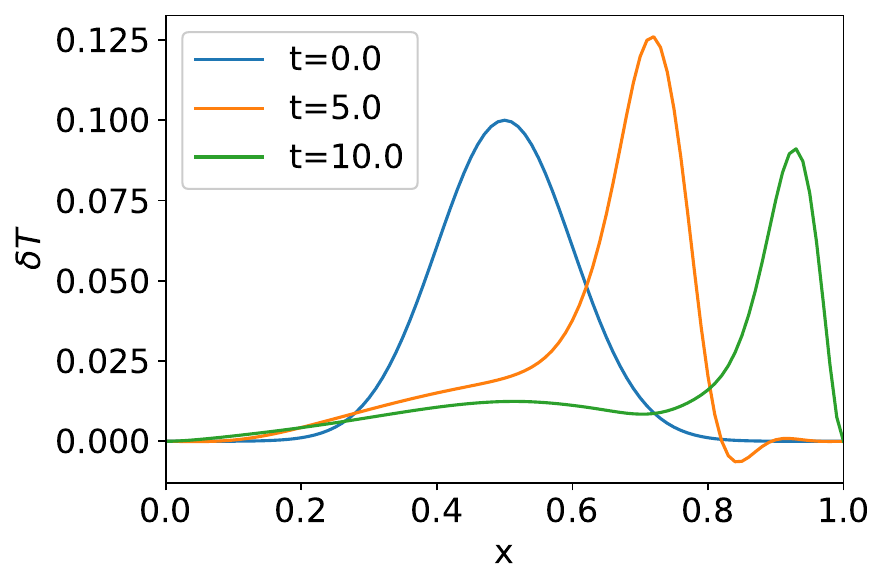}
\caption{Jam dynamics of a single pulse. An isolated pulse initially grows via jamming instability with negative conductivity, and then saturates.}
\label{Fig:Single}
\end{figure}

\begin{figure}[!h]
\centering\includegraphics[width=0.5\textwidth]{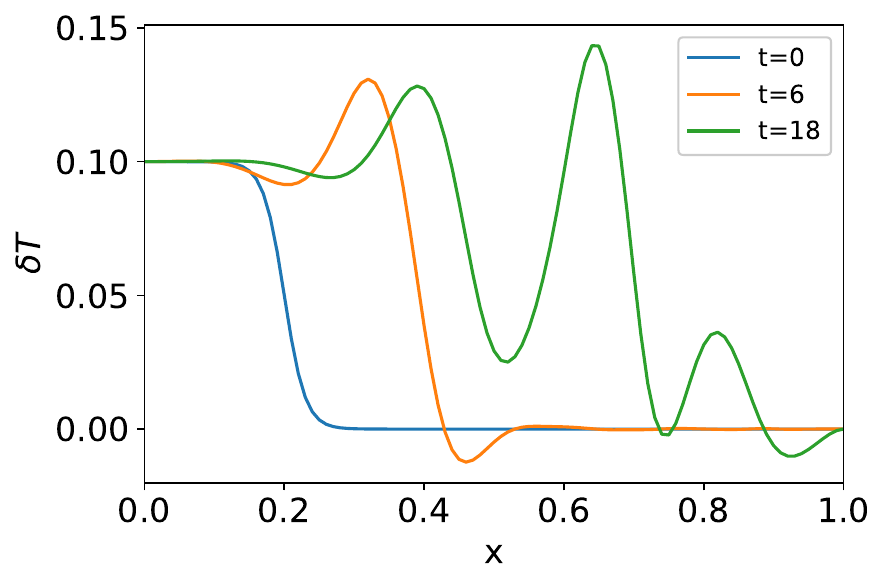}
\caption{Formation of multiple trains of jams, jamiton}
\label{Fig:Multiple}
\end{figure}

\begin{figure}[!h]
\centering\includegraphics[width=0.5\textwidth]{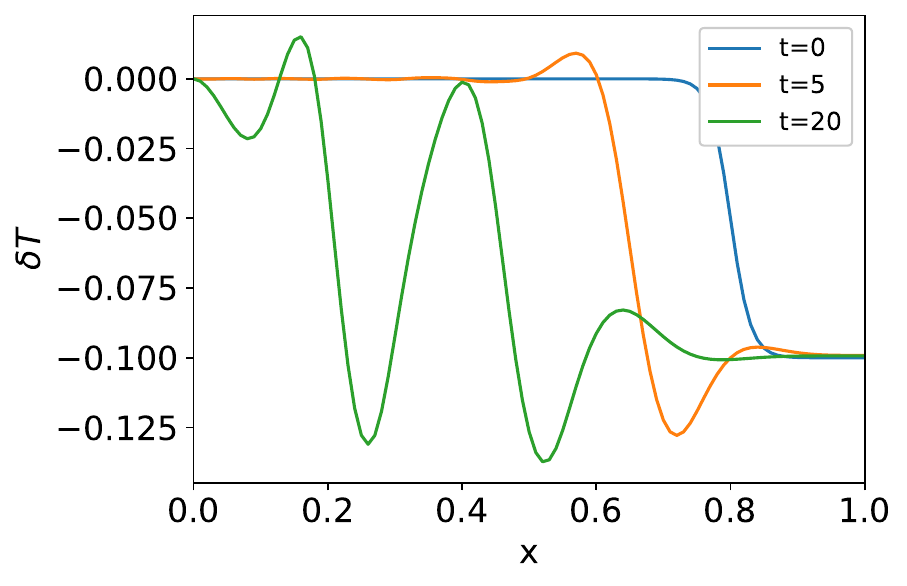}
\caption{Jamiton can form for a cold front propagating inward.}
\label{Fig:MultipleCold}
\end{figure}

Jam formation and saturation are illustrated by numerically solving Eq.(\ref{Eq:NLtelegraph}). Since jamming is triggered by negative conductivity, additional damping is required to regulate its growth. While hyperconductivity is a candidate, we also include the suppression and regulation by shear flow, as
\begin{align}
\tau_d\frac{\partial^2}{\partial t^2}\delta T+\frac{\partial}{\partial t}\delta T+\frac{\alpha}{1+\beta(\partial_x^2\delta T)^2}\delta T\partial_x\delta T-\chi_2\partial_x^2\delta T+\chi_4\partial_x^4\delta T=\tilde{s}
\end{align}
Here the shear flow suppression is included in the nonlinear term. The curvature of the temperature perturbation mimics the shear flow suppression, since $v_\theta\propto E_r\propto\partial_x\delta T$ so $v_\theta'\propto\partial_x^2\delta T$. We note that the $d\delta Q/d\delta T<0$ condition is not met for this choice (i.e. $n=1$ here). Thus growth is entirely due to the time delay. Fig.(\ref{Fig:Single}) illustrates the dynamics of a single pulse. The pulse initially grows via negative conductivity, and then saturates. Interestingly, a train of multiple jams can form, as shown in Fig.(\ref{Fig:Multiple}). In this case, the first jam forms (at $t=6$) and then propagates forward. The second jam forms and follows the first jam. This process continues to form multiple layers of jams. In this simulation, 3 jams are observed at $t=18$. Multiple layers of jams can form from a propagating cold front, as shown in Fig.\ref{Fig:MultipleCold}. It is interesting to note that in both cases, the emergence of the pattern is progressive, unlike the formation of a spatial eigenmode.



The ultimate aim of this work is to explore jams as a mechanism for layering. The discussion so far has focused on jams as transport barriers, jam formation and pattern structure. Here, we discuss jams and jamiton trains, in particular, as a route to layering and staircase formation.

Recall that a staircase self-assembles from an array of layers or transport barriers, which may be viewed as phase domain boundaries. A staircase, then is a quasi-periodic array of transport barriers. A staircase temperature or density profile consists of an array of 'steps' with local gradient shallower than the mean gradient, interspaced by 'jumps' with locally steep gradients. The jumps correspond to local transport barriers, and frequently are supported by $E\times B$ shear layers. Staircase profiles can enjoy improved global confinement.

A jamiton train, which may be thought of as an array of temperature profile corrugations, is thus a natural candidate for the formation of layers. Fig.\ref{Fig:Multiple} shows a jamiton train - i.e. an array of sharply localized corrugations (i.e. $\delta T(x)$ may be thought of as corrugation on the mean profile $T_0(x)$). From this perspective, $\Delta^2\sim k_{max}^{-2}$ is the width of the jamp-to-jump spacing (i.e. 'step size'). The $E\times B$ sear associated with the barrier is $v'_E\sim(c/eB)\delta T''$ (using radial force balance) so
\begin{equation}
v'_E\sim\frac{c\delta T}{eB\Delta_{max}^2}\sim\frac{\omega_{ci}\rho_i^2\alpha T_i}{2(\chi_2\chi_4)^{1/2}}\left(\frac{\delta T}{T}\right)^2
\end{equation}
Balancing shearing rate and modulation growth rate $\gamma_{max}$ gives
\begin{subequations}
\begin{equation}
\frac{\delta T}{T}\sim\left(\frac{1}{v_{thi}\rho_i}\right)\left(\frac{\chi_4}{\tau_R}\right)^{1/2}
\end{equation}
as the corrugation amplitude, and
\begin{equation}
\Delta_{max}^2\sim\rho_i^2\frac{2v_{thi}}{\alpha T_i}\sqrt{\frac{\chi_2\tau_R}{\rho_i^2}}
\end{equation}
as the step width scale. Using standard parameter values gives plausible estimates for $\Delta_{max}$ so $\rho_i<\Delta_{max}<L_T$, where $L_T$ is the mean profile scale length. A jamiton and the resulting corrugated temperature profile are shown in Figs. (\ref{Fig:Multiple}) and (\ref{Fig:Staircase}).

\begin{figure}[!h]
\centering\includegraphics[width=0.5\textwidth]{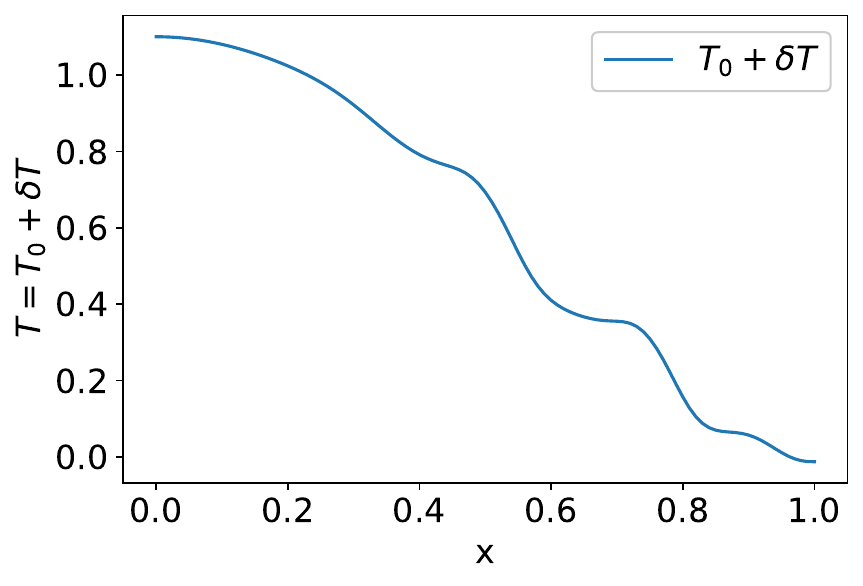}
\caption{Temperature staircase. $T_{tot}=T_0+\delta T$, where $T_0=(1-x^2)^2$ and $\delta T$ from Fig.{\ref{Fig:Multiple}}.}
\label{Fig:Staircase}
\end{figure}
\end{subequations}

\section{Particle Flux Avalanching in Dissipative Drift Turbulence}

The main application up to this point was heat transport. However, the methodology applies to particle transport as well. Putting the two together
in a coherent manner would require confronting the issue of coexistence and the competition of the mechanisms of
double diffusion and staircase formation through jamming in two separate
channels. As a first approach, we can consider the two as being separate and basically 
replace heat flux with particle flux since the two have the same structure. Here we start by considering a simple model of dissipative drift wave
turbulence, the Hasegawa-Wakatani model \cite{Hasegawa}, which we take as a minimal description of instability driven ``beta plane'' turbulence, that has waves and eddys, conserves potential vorticity, and can form staircases. As is commonly the case in such models,
it can be formulated either as a fixed gradient, or a flux driven system, in this case
using the penalisation method thanks to the recently developed P-FLARE
code\cite{Guillon25ar}. Both of these approaches are useful for understanding different
aspects of the system, even though the flux driven formulation is
clearly more relevant for the underlying physics involved.

\begin{figure}
\begin{centering}
\includegraphics[width=0.5\textwidth]{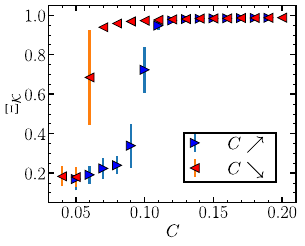}
\par\end{centering}

\caption{
  The hysteresis loop that is observed in the evolution of the zonal kinetic energy fraction $\Xi_K$ as a function of the control parameter $C/\kappa$, which is very slowly increased (blue), and then decreased (red) in fixed gradient Hasegawa-Wakatani simulations.}
\label{fig:loop}
\end{figure}

In particular, considering the Hasegawa-Wakatani system written in
dimensionless, fixed gradient form (i.e. $e\Phi/T\rightarrow\Phi$,
$\delta n/n_{0}\rightarrow n$, $x/\rho_{s}\rightarrow x$, $y/\rho_{s}\rightarrow y$
and $\Omega_{i}t\rightarrow t$, and $\kappa\equiv-\rho_{s}n_{0}^{-1}dn_{0}/dx$,
and $C\equiv v_{te}^{2}k_{\parallel}^{2}/\left(\nu_{c}\Omega_{i}\right)$,
where $\nu_{c}$ and $\Omega_{i}=eB/mc$ are the collision and the
ion cyclotron frequency respectively, $\rho_{s}=c_{s}/\Omega_{i}$
is the so called ion sound Larmor radius with $c_{s}=\sqrt{T/m}$
and $v_{te}=\sqrt{T/m_{e}}$ being the plasma sound speed and the
electron thermal velocities respectively), we have:
\begin{equation}
\left(\partial_{t}+\hat{\mathbf{b}}\times\nabla\Phi\cdot\nabla\right)n+\kappa\partial_{y}\widetilde{\Phi}=C\left(\widetilde{\Phi}-\widetilde{n}\right)\label{eq:hwd}
\end{equation}
\begin{equation}
\left(\partial_{t}+\hat{\mathbf{b}}\times\nabla\Phi\cdot\nabla\right)\zeta=C\left(\widetilde{\Phi}-\widetilde{n}\right)\label{eq:hwv}
\end{equation}
Dividing (\ref{eq:hwd}) and (\ref{eq:hwv}) by $\kappa^{2}$, and
normalizing both the field amplitudes and $t^{-1}$ by $\kappa$,
we can show the inviscid dynamics to depend only on the ratio $C/\kappa$.
Since the actual dissipation terms are quite small, adding viscous
dissipation on both equations (as small as necessary to remove the
small scale fluctuations) does not alter this observation.

\begin{figure}
\begin{centering}
\includegraphics[width=0.8\textwidth]{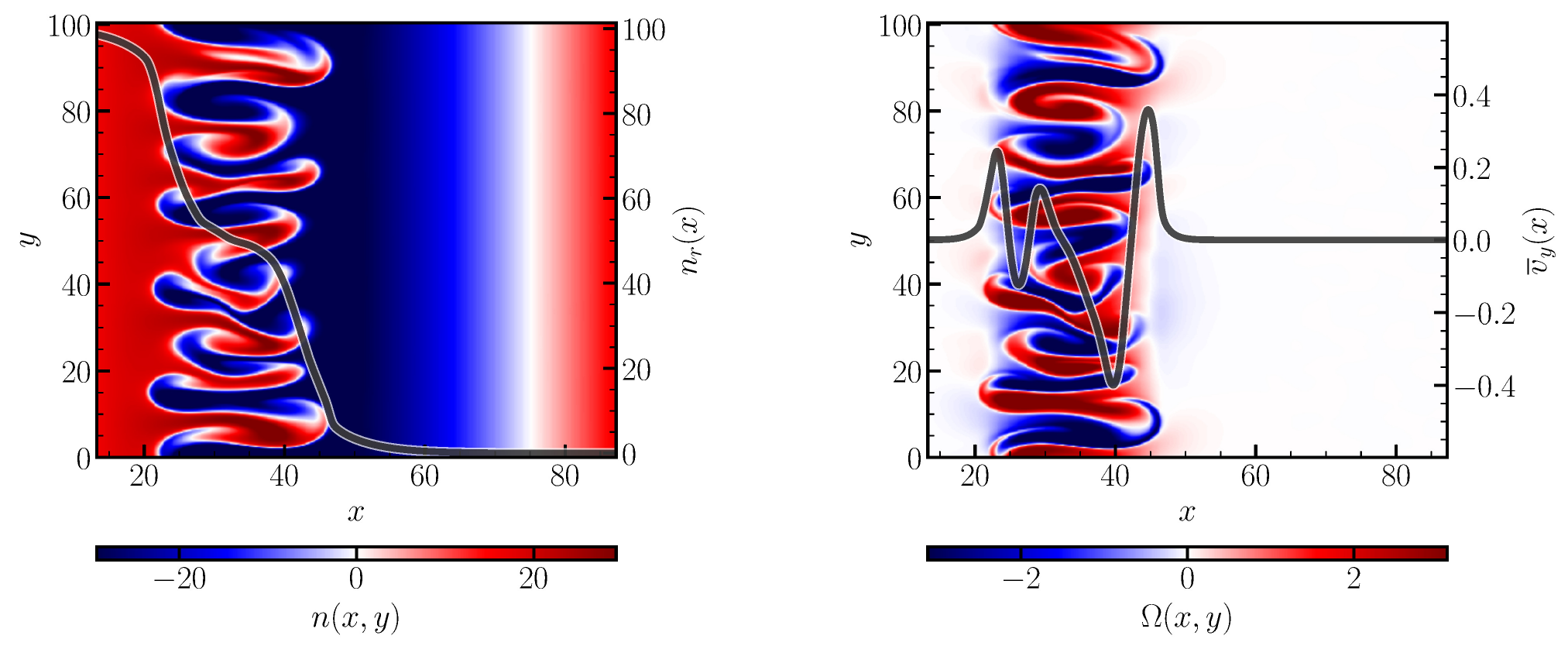}
\par\end{centering}

\caption{Blobs and voids moving through each other in a fingering instability
  pattern in density (left), and through asymmetric dipole structures in vorticity (right).}
\label{fig:hwblobs}
\end{figure}

Another interesting aspect of the system (\ref{eq:hwd}-\ref{eq:hwv}),
sometimes called the modified Hasegawa-Wakatani model, is that even
though in its inviscid form, it has no stability threshold, it is
well known to drive zonal flows if its control parameter $C/\kappa$
is greater than a threshold value of about $0.1$. Below that, one finds
basically two dimensional turbulence, dominated by eddies.
Therefore while it has no linear threshold for instability,
it manifests a nonlinear threshold, and as such, it is an interesting
example for the application of the above phenomonology, which also applies
for a nonlinear, potentially self organized bistable (SOB) state \cite{diSanto16}.
Note that from a thermodynamical perspective the ``control'' parameter $C/\kappa$ plays the role of the inverse
``temperature'', such that large values of $C/\kappa$ results in
a ``cold'', organized state, while smaller values of $C/\kappa$
results in a ``hot'' disordered state.

Using the ratio of zonal to total kinetic energy as the order parameter
$\Xi_{K}=K_{Z}/K$, the transition controlled by $C/\kappa$ appears
to be an abrupt transition, with features that are similar to a first
order phase transition. In particular, it displays a clear hysteresis loop pattern\cite{Guillon25} as can be seen in figure \ref{fig:loop}. This suggests that once formed, it takes more energy to melt the staircase structure \cite{Ramirez} than to form it at first.
In some sense this can be seen as a phase transition from 2D
to quasi-1D turbulence, in the same vein as the transition from 3D
to 2D turbulence as one increases rotation rate, or decreases the
fluid depth\cite{Benavides}. 

However while the fixed gradient model is useful for understanding scaling characteristics of the system, in order to describe the jamming nature of the transition 
one has to use the flux driven formulation. There one defines an initial density profile, and a source term in the density equation (which can be zero if we are interested in the relaxation problem), and the gradient $\kappa$ adapts itself to this imposed steady state flux. In order to see that the above one dimensional formulation applies to flux driven Hasegawa-Wakatani model, we first start with an initial condition and let the system relax. We observe that the sharp gradient region gets flattened by generating a ``blob'' that moves outward and a ``void'' that moves inward. Looking in more detail at the 2D evolution, it is actually
a train of blobs and voids that are moving radially outward and inward while passing through each-other as can be seen in figure \ref{fig:hwblobs}. These seem to be supported by asymmetric dipolar structures with nonzero net vorticity. In this sense, it is clear that the underlying dynamics of the flux-driven Hasegawa-Wakatani respects JRS, and therefore, can help us better understand the details and the meaning of the one dimensional equations that we use to model the avalanche transport.

\begin{figure}
\begin{centering}
\includegraphics[width=0.8\textwidth]{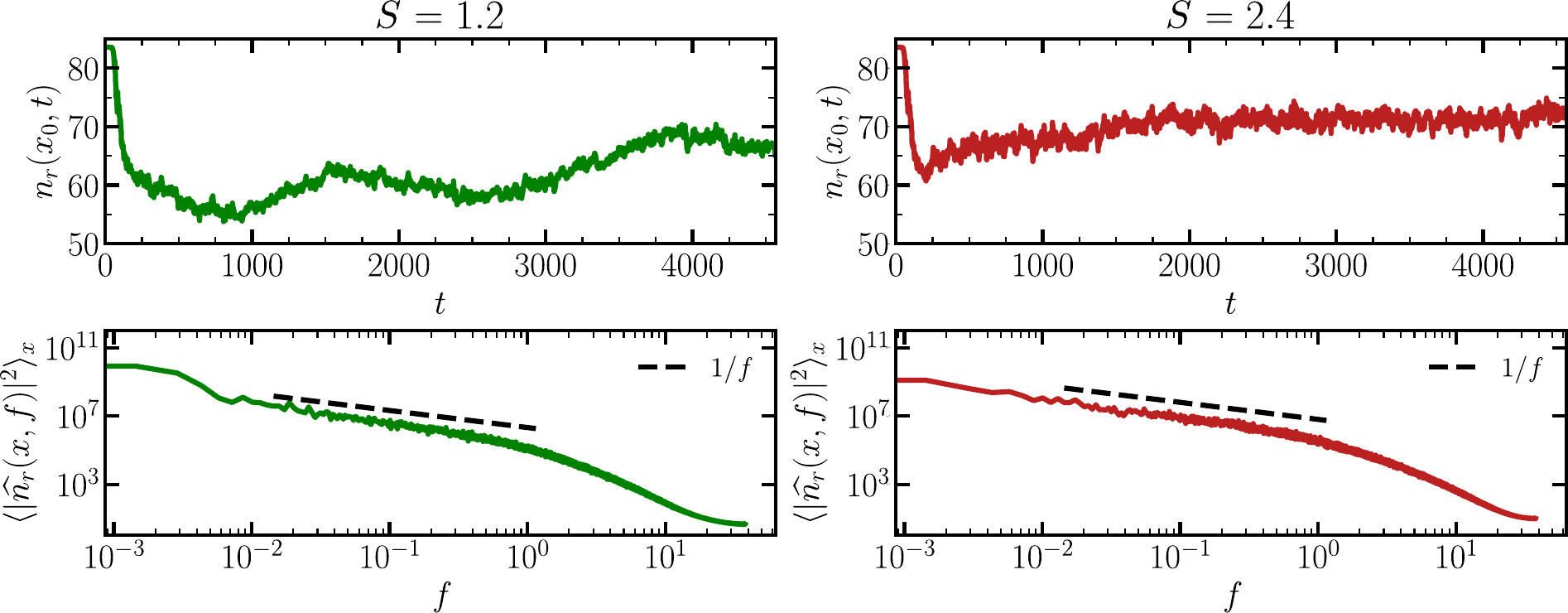}
\par\end{centering}

\caption{Top row: time series of the density profile at the source center, corresponding
to sandpile height, observed in the evolution of two different source levels $S$,
both of which are close to the marginally stable region for
the eddy dominated to zonal flow dominated transition. Even though
the drive and the zonal flow fractions are different for the two simulations,
both have very similar density profiles, and both clearly show $1/f$
scaling of the frequency spectrum (bottom row).}
\end{figure}\label{fig:freqspec}

In the flux driven case, near marginality for the eddy-dominated to zonal flow-dominated transition one observes features of critical behavior with a clear $1/f$ frequency spectrum, as can be seen in figure \ref{fig:freqspec}. Note that even though the Hasegawa-Wakatani system, as opposed to, for instance, the interchange, does not favor ``avalanches'' (i.e. radially extended convective cells, or ``streamers'') linearly, the coupled evolution of the density profile together with a self-consistent flux in the parameter regime which is critical to staircase formation, effectively reproduces avalanching statistics implied by the criticality phenomenology.

\begin{figure}
\begin{centering}
\includegraphics[width=0.4\textwidth]{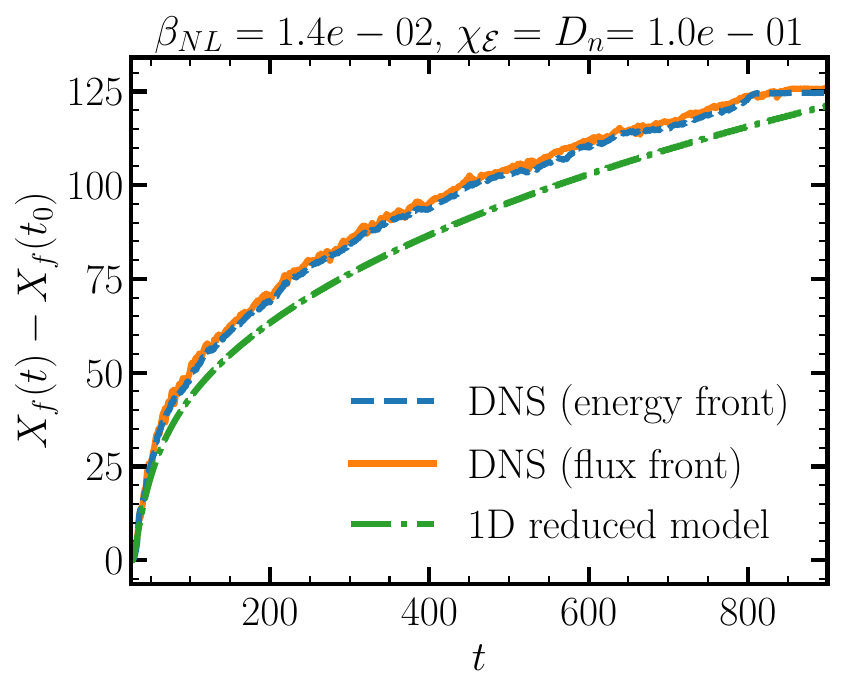}
\par\end{centering}
\caption{The comparison between position of the relaxation front
of density in flux driven Hasegawa-Wakatani simulations, defined using the energy (blue) or the particle flux (orange) profiles, and the 1D model (green). Parameters used for the 1D model are given above the figure.}\label{fig:1dcomp}
\end{figure}

Second, we show that spreading of the flux as discussed in Eqn. \ref{Eq:DelayPhasetrophy}, but retaining the spreading term $\partial_x\Gamma_Q(\mathcal{E})$, coupled with an equation for the density profile $n_r$, can be used to match the relaxation dynamics rather accurately, as can be seen in figure \ref{fig:1dcomp}. Indeed, during the relaxation of an initially steep density profile, such as the one shown in figure \ref{fig:hwblobs}, one can simultaneously observe the spreading of the turbulent front towards the unperturbed flat profile region. Such a front can be defined either using the radial profile $\mathcal{E}(x)$ of the energy (blue), or that of the particle flux $\Gamma_n(x)$ (orange), and both definitions match, as shown in figure \ref{fig:1dcomp}. Using the following 1D model for the coupled density relaxation/energy spreading
\begin{equation}
\partial_{t}\mathcal{E}+\partial_x\Gamma_\mathcal{E}(\mathcal{E})=2\gamma_{max}\mathcal{E}-\beta_{NL}\mathcal{E}^{2}\,,\label{Eq:1DTKE}
\end{equation}
\begin{equation}
\partial_{t}n_{r}+\partial_x\Gamma_n(\mathcal{E})=0\,,\label{Eq:1Dnr}
\end{equation} 
where $\gamma_{max}$ is the growth rate of the linear instability, $\beta_{NL}$ is eddy-damping non-linear saturation, and $\Gamma_\mathcal{E}(\mathcal{E})$ and $\Gamma_n(\mathcal{E})$ are respectively the spreading and particle relaxation fluxes. Notice that the first equation could be equally written on the particle flux $\Gamma_n$, similarly to Eqn. \ref{Eq:DelayPhasetrophy}, but using the energy is easier if one wants to include the linear growth rate and is more convenient in order to solve a model in practice. To model these terms, we use the classic spreading expressions $\Gamma_\mathcal{E}=-\chi_\mathcal{E}\mathcal{E}\partial_x\mathcal{E}$ and $\Gamma_n=-D_n\mathcal{E}\partial_xn$. These expressions are well-known to be too simple, and particularly miss out bistability. Nevertheless, in this particular example, they seem to hold using appropriate diffusion coefficients $\chi_\mathcal{E}$, $D_n$, and reproduce the flux-driven DNS, as shown in Figure \ref{fig:1dcomp} (green).

Note however that imposing a greater flux, would eventually push the system away from the staircase formation region and result in a two dimensional turbulence state dominated by eddies. Even though the interplay of heat and particle fluxes is beyond the scope here, in a coupled density/temperature system, it is possible that the flux gradient relations become intrinsically coupled, and increasing the particle flux, may not result in a full collapse of the staircase as both the temperature and the density gradients would respond, and may work against each-other.

In short, using a simple flux driven formulation of the Hasegawa-Wakatani system, we find that it respects JRS, with blobs and voids moving in opposite directions, and a detailed relaxation behavior consistent with a 1D model of spreading as described in this paper. When driven slowly so that it saturates to at criticality for the phase transition between zonal flow to eddy dominated states, the model displays critical behavior, including profile stiffness and $1/f$ avalanche statistics.





\section{Discussion and Conclusion}

This paper is motivated by understanding layering and staircase formation in avalanching drift wave turbulence. Special focus is on the topics of transport barrier formation - as a jamming process - and on layering via jam arrays. We exploit intuition gained from hydrodynamic models of traffic flow and from MIPS phenomena. The 'base state' here is one of avalanching turbulence, i.e. we take the temperature to avalanche, so there exists a field of dynamic temperature pulses or corrugations $\delta T(x,t)$ which propagate along a more slowly evolving 'background' profile. The crux of the matter is to understand how heat flux jams form.

The upshot is that the constitutive relation $Q[\delta T]$, which relates the heat flux to the avalanche field is the key determinant of jam formation. $Q[\delta T]$ may be thought of as a generalization of the familiar flux-gradient relation $Q=-\chi\nabla T$, familiar from kinetic theory. In the case where $Q[\delta T]$ is given explicitly, i.e. $Q=f(\delta T)$, we show jamming occurs for $dQ/d\delta T<0$ so that the flux \textit{decreases} with $\delta T$, for some range of $\delta T$. This is analogous to the well known kinematic wave model result which requires $d(\rho V(\rho))/d\rho<0$, so that waves propagate backwards against the flow and ultimately form backward shocks - i.e. jams. Backward shocks are frequently encountered at toll booths and other bottlenecks. We also demonstrated the compatibility of forms of $Q$ allowing jamming with the principle of joint reflection symmetry, which constrains the structure of $Q[\delta T]$. The $dQ[\delta T]/d\delta T<0$ condition is also connected to the oft-invoked condition of bistability for barrier onset, and to the theory of MIPS. The condition for triggering barrier onset, i.e. $\delta T>\delta T_{crit}$, is calculated. $\delta T_{crit}$ is set by the size of the avalanche required to access the $dQ/d\delta T<0$ branch of the $Q[\delta T]$ curve. $\delta T_{crit}$ may be thought of as an instantaneous threshold for jam formation. Several aspects of jams are discussed in detail.

In avalanching turbulence, the use of (what amounts to) a mean field model for $Q[\delta T]$ is dubious. A more plausible constitutive relation is one of the Guyer-Krumhansal form of $\partial_tQ=-(Q-Q_0)/\tau_R$, where $Q$ relaxes to a form $Q_0$ (of the mean field variety) on the time scale $\tau_R$. The response time $\tau_R$ is due to the mixing of phase space intensity $\langle\delta f^2\rangle$ by turbulent scattering, and is linked to the fundamental structure of the gyrokinetic equation. The appearance of $\tau_R$ defines another speed in the problem - $(\chi_{neo}/\tau_R)^{1/2}$. Retaining finite relaxation converts the familiar Burgers-type equation for $\delta T$ to a nonlinear telegraph equation. Crucially, jams will occur in this system for sufficiently long $\tau_R$, i.e. $\chi_{neo}/\tau_R<c_0^2\sim(\alpha^2\delta T_0^2)$. Note that "how long is 'long'?" is determined by the avalanche size (bigger is better!) and the neoclassical heat diffusivity (smaller favors jams, as expected). Crucially, bistability of $Q_0$- or, equivalently, $dQ_0/d\delta T<0$ - is \textit{not} required for jam formation! Thus, we see that jam formation is a more general phenomenon than bifurcations between states of a bistable flux. This is an important observation!

As regimes of long flux response time are seen to be conducive to jam formation, it becomes necessary to characterize these. Now, since $\tau_R$ is determined by turbulent mixing of $\langle\tilde{f}(1)\tilde{f}(2)\rangle$, we have $1/\tau_R\sim1/\tau_c$, the turbulent correlation time. As $1/\tau_c\sim k_\perp^2 D(\mathcal{E})$, where $\mathcal{E}$ is turbulence intensity, we expect long $\tau_c$, and thus long $\tau_R$, to occur in regimes close to marginality, where intensity $\mathcal{E}$ is weak. Thus, we ansatz $1/\tau_R\sim(R/L_T-R/L_{T_{crit}})^{\mu}/\tau_0$, where the critical scaling exponent $\mu$ is related to the gradient stiffness exponent, and $\tau_0$ is a 'typical' value of $\tau_c$, a scale factor. Then the jam formation condition is $c_0^2\tau_0/\chi_{neo}>(R/L_T-R/L_{T_{crit}})^{\mu}$, and this condition determines 'how near is "near"' to marginal stability\cite{SOCMarginal}. Gyrokinetic simulations have examined the dependence of $\tau_R$ on proximity to marginality, and noted $\tau_R$ is large in that limit. See Fig.\ref{Fig:ResponseTime}. Typical scales and amplitude of a nonlinear telegraph equation wave train were calculated. Numerical solutions of the nonlinear telegraph equation manifest a train of corrugations $\delta T$. Super-posing these on a smooth background profile gives a staircase structure.

We also presented the results of a simple study of a flux driven Hasegawa-Wakatani system. Consistent with JRS, a fundamental constituent of the theory presented here, avalanching is manifested as a two trains, of outgoing blobs and incoming voids, respectively. The relaxation behavior of this simple - but - realistic system is similar to that of the 1D model discussed at length here. Studies show that turbulence spreading plays a key role in the pulse propagation dynamics discussed in the bulk in this paper.

Many questions remain. Direct simulation of jamming is still in its early stages. Analysis of jamming criteria, and especially their sensitivities to deviation from criticality, marginality, etc. remain incomplete. Macroscopic characterization of jamiton train - induced staircase formation is terra incognita. Since it seems clear that the study of the physics of heat flux jams has benefited greatly from the many analogous results obtained from the study of traffic flow, it is perhaps worthwhile to conclude with a speculation in this vein. A recurring question concerning layering in confined plasma is : "yes- it is interesting, but how does it affect confinement? Does it 'matter'?" Here, it is interesting to refer to yet another traffic flow analogy, namely that of the states of flow through an array of traffic lights. As we know from everyday life, the state of flow will depend on the ratio of time green $\tau_g$ to time red $\tau_r$. Each light, then, can be viewed as a permeable barrier - a 'jump' in a staircase with permeability $p$ related to the ratio $\tau_g/\tau_r$. Interestingly, the overall state of the flow has been shown to be 'free flowing' for $\tau_g/\tau_r>p_{crit}$, and a 'crawl' for $\tau_g/\tau_r<p_{crit}$. $p_{crit}$ is determined by the flux density of traffic $\rho V$ and the maximal $\rho V$. Of course, a crawl state is analogous to a state of good confinement, i.e. low flow or low turbulent transport. So it is tempting to speculate that an array of time varying barriers (i.e. driven shear layers) could be tunable (via $\tau_g/\tau_r$)  to produce both free flow transport and good confinement (crawl) states, without a steady state of strongly suppressed turbulence. This is an interesting topic for future investigation.

\begin{acknowledgments}
We thank M. Vergassola, M. Choi, T. Long, F. Kin, G. Dif-Pradalier for useful discussions. P.D. acknowledges a discussion with M. Cates concerning the kinematic wave - MIPS duality. The authors thank participants in the 2024 INI program "Antidiffusion: from Sub-Cellular to Astrophysical Scales" and the participants in the 2022 and 2025 Festival de Theorie for stimulating interactions. This research was in part supported by the U.S. Department
of Energy, Office of Science, Office of Fusion Energy Sciences under Award No. DE-FG02-04ER54738; the Sci DAC ABOUND Project, scw1832; the EPSRC under Grant No. EP/R014604/1; the Grants-in-Aid for Scientific Research of JSPS of Japan (JP21H010166,JP23K20838); the Jean Zay supercomputer of IDRIS under the allocation AD010514291R2 by GENCI; the European Union via the Euratom Research and Training Programme (Grant Agreement No. 101052200—EUROfusion); the framework of the French Research Federation for Fusion Studies.
\end{acknowledgments}


\end{document}